\documentclass[prl,twocolumn,longbibliography,
superscriptaddress,
 amsmath,amssymb,
 aps,floatfix
]{revtex4-1}
\usepackage{gensymb}
\usepackage{graphicx}% Include figure files
\usepackage{dcolumn}% Align table columns on decimal point
\usepackage{bm}% bold math
\usepackage{multirow}
\usepackage{textcomp, gensymb}
\usepackage{
}
\usepackage{graphicx}
\usepackage{dcolumn}
\usepackage{bm}
\usepackage{amssymb, amsmath}
\usepackage{microtype}
\usepackage{xfrac}
\usepackage{array}
\usepackage{gensymb}
\usepackage{xcolor}
\usepackage[normalem]{ulem}
\usepackage{dsfont}
\usepackage{diagbox}

\usepackage[T1]{fontenc}

\usepackage{indentfirst} 
\usepackage{float}
\usepackage{graphicx}
\usepackage{ulem}
\usepackage{setspace}
\usepackage{physics}

\newcommand{\fgt}{Fe$_5$GeTe$_2$}

\newcommand{\ef}{\textrm{$E_F$}}
\newcommand{\kf}{$k_F$}

\newcommand{\tc}{$T_\mathrm{C}$}

\begin{document}
 
%\title{Intertwined flat bands and sublattice ordering in a near room temperature van der Waals ferromagnet}
\title{Dichotomy of flat bands in the van der Waals ferromagnet Fe$_5$GeTe$_2$}

\author{Han Wu}
\affiliation{Department of Physics and Astronomy and Rice Center for Quantum Materials, Rice University, Houston, TX, 77005 USA}

\author{Jianwei Huang}
\affiliation{Department of Physics and Astronomy and Rice Center for Quantum Materials, Rice University, Houston, TX, 77005 USA}

\author{Chaowei Hu}
\affiliation{Department of Physics, University of Washington, Seattle, Washington 98195, USA}
\affiliation{Department of Materials Science and Engineering, University of Washington, Seattle,
Washington 98195, USA}

\author{Lei Chen}
\affiliation{Department of Physics and Astronomy and Rice Center for Quantum Materials, Rice University, Houston, TX, 77005 USA}

\author{Yiqing Hao}
\affiliation{Neutron Scattering Division, Oak Ridge National Laboratory, Oak Ridge, TN 37831, USA}

\author{Yue Shi}
\affiliation{Department of Physics, University of Washington, Seattle, Washington 98195, USA}

\author{Paul Malinowski}
\affiliation{Department of Physics, University of Washington, Seattle, Washington 98195, USA}

\author{Yucheng Guo}
\affiliation{Department of Physics and Astronomy and Rice Center for Quantum Materials, Rice University, Houston, TX, 77005 USA}

\author{Bo Gyu Jang}
\affiliation{Theoretical Division and Center for Integrated Nanotechnologies, Los Alamos National Laboratory, Los Alamos, NM, USA}
\affiliation{Department of Advanced Materials Engineering for Information and Electronics, Kyung Hee University, Yongin 17104, Republic of Korea}

\author{Jian-Xin Zhu}
\affiliation{Theoretical Division and Center for Integrated Nanotechnologies, Los Alamos National Laboratory, Los Alamos, NM, USA}

%\author{Fang Xie}
%\affiliation{Department of Physics and Astronomy and Rice Center for Quantum Materials, Rice University, Houston, TX, 77005 USA}

\author{Andrew F. May}
\affiliation{Materials Science and Technology Division, Oak Ridge National Laboratory, Oak Ridge, TN 37831}

\author{Tyler Werner}
\affiliation{Department of Applied Physics, Yale University, New Haven, Connecticut 06511, USA}

\author{Siqi Wang}
\affiliation{Department of Applied Physics, Yale University, New Haven, Connecticut 06511, USA}

\author{Xiang Chen}
\affiliation{Department of Physics, University of California, Berkeley, Berkeley, California 94720, USA}

%\author{Chandan Setty}
%\affiliation{Department of Physics and Astronomy and Rice Center for Quantum Materials, Rice University, Houston, TX, 77005 USA}

\author{Yaofeng Xie}
\affiliation{Department of Physics and Astronomy and Rice Center for Quantum Materials, Rice University, Houston, TX, 77005 USA}

\author{Bin Gao}
\affiliation{Department of Physics and Astronomy and Rice Center for Quantum Materials, Rice University, Houston, TX, 77005 USA}

\author{Yichen Zhang}
\affiliation{Department of Physics and Astronomy and Rice Center for Quantum Materials, Rice University, Houston, TX, 77005 USA}

\author{Ziqin Yue}
\affiliation{Department of Physics and Astronomy and Rice Center for Quantum Materials, Rice University, Houston, TX, 77005 USA}

\author{Zheng Ren}
\affiliation{Department of Physics and Astronomy and Rice Center for Quantum Materials, Rice University, Houston, TX, 77005 USA}

\author{Makoto Hashimoto}
\affiliation{Stanford Synchrotron Radiation Lightsource, SLAC National Accelerator Laboratory, Menlo Park, California 94025, USA}

\author{Donghui Lu}
\affiliation{Stanford Synchrotron Radiation Lightsource, SLAC National Accelerator Laboratory, Menlo Park, California 94025, USA}

\author{Alexei Fedorov}
\affiliation{Advanced Light Source, Lawrence Berkeley National Laboratory, Berkeley, CA 94720, USA}

\author{Sung-Kwan Mo}
\affiliation{Advanced Light Source, Lawrence Berkeley National Laboratory, Berkeley, CA 94720, USA}

\author{Junichiro Kono}
\affiliation{Department of Materials Science and NanoEngineering, Rice University, Houston, TX, 77005, USA}
\affiliation{Departments of Electrical and Computer Engineering, Rice University, Houston, TX, 77005, USA}
\affiliation{Department of Physics and Astronomy and Rice Center for Quantum Materials, Rice University, Houston, TX, 77005 USA}

\author{Yu He}
\affiliation{Department of Applied Physics, Yale University, New Haven, Connecticut 06511, USA}

\author{Robert J. Birgeneau}
\affiliation{Department of Physics, University of California, Berkeley, Berkeley, California 94720, USA}
\affiliation{Materials Sciences Division, Lawrence Berkeley National Laboratory, Berkeley, California 94720, USA}
\affiliation{Department of Materials Science and Engineering, University of California, Berkeley, USA}

\author{Pengcheng Dai}
\affiliation{Department of Physics and Astronomy and Rice Center for Quantum Materials, Rice University, Houston, TX, 77005 USA}

\author{Xiaodong Xu}
\affiliation{Department of Physics, University of Washington, Seattle, Washington 98195, USA}
\affiliation{Department of Materials Science and Engineering, University of Washington, Seattle,
Washington 98195, USA}

\author{Huibo Cao}
\affiliation{Neutron Scattering Division, Oak Ridge National Laboratory, Oak Ridge, TN 37831, USA}

\author{Qimiao Si}
%\email{qmsi@rice.edu}
\affiliation{Department of Physics and Astronomy and Rice Center for Quantum Materials, Rice University, Houston, TX, 77005 USA}

\author{Jiun-Haw Chu}
%\email{jhchu@uw.edu}
\affiliation{Department of Physics, University of Washington, Seattle, Washington 98195, USA}

\author{Ming Yi}
\email{mingyi@rice.edu}
\affiliation{Department of Physics and Astronomy and Rice Center for Quantum Materials, Rice University, Houston, TX, 77005 USA}

\date{\today}

\begin{abstract}

Quantum materials with bands of narrow bandwidth near the Fermi level represent a promising platform for exploring a diverse range of fascinating physical phenomena, as the high density of states within the small energy window often enables the emergence of many-body physics. On one hand, flat bands can arise from strong Coulomb interactions that localize atomic orbitals. On the other hand, quantum destructive interference can quench the electronic kinetic energy. Although both have a narrow bandwidth, the two types of flat bands should exhibit very distinct spectral properties arising from their distinctive origins. So far, the two types of flat bands have only been realized in very different material settings and chemical environments, preventing a direct comparison. Here, we report the observation of the two types of flat bands within the same material system--an above-room-temperature van der Waals ferromagnet, Fe$_{5-x}$GeTe$_2$, distinguishable by a switchable iron site order. The contrasting nature of the flat bands is also identified by the remarkably distinctive temperature-evolution of the spectral features, indicating that one arises from electron correlations in the Fe(1) site-disordered phase, while the other geometrical frustration in the Fe(1) site-ordered phase. Our results therefore provide a direct juxtaposition of the distinct formation mechanism of flat bands in quantum materials, and an avenue for understanding the distinctive roles flat bands play in the presence of magnetism, topology, and lattice geometrical frustration, utilizing sublattice ordering as a key control parameter.

%While near Fermi level flat bands can exhibit diverse emergent behaviors with the enhancement of thermal fluctuations, experimental investigations into these different flat band behaviors have been predominantly limited to distinct materials. An outstanding question in this context is how the temperature evolution of flat bands is controlled. To address this and explore near Fermi surface physics in a narrow-electronic-band system, there is a need to engineer flat bands within a single system. Here, we report the switchable distinct temperature dependent flat bands evolution near Fermi level in a 3-d near room temperature van der Waals (vdW) ferromagnet Fe$_{5-x}$GeTe$_2$, arising from tunable Fe(1) sites occupancy. Furthermore, we showcase a coherence-incoherence crossover is intertwined with the Fe(1) site-order. The distinct flat bands behaviors are potentially related to the difference origin of the flat bands that correlations induces flat bands in the Fe(1) site-disordered phase and geometrical frustration enforced flat bands in Fe(1) site-ordered phase. Beyond uncovering the distinct temperature evolution behaviors of flat bands near the Fermi level, our results provide a potential avenue for understanding different correlated flat band behaviors in the presence of magnetism, topology, and lattice geometrical frustration, utilizing sublattice tuning as a key control parameter.
\end{abstract}

\maketitle
%%%%%%%%%%%%%%%%%%%%%%%%%%%%%%%%%%%%%%%%%%%%%%%%%%

\clearpage

\section{Introduction}
In systems characterized by narrow electronic bands or flat bands, the kinetic energy of electrons ($t$) is notably suppressed compared to the energy scale of electron correlations ($U$). Such system can become a versatile platform for exploring various correlated electronic phases~\cite{Calugaru2022,Regnault2022,Checkelsky2024-yn,Bernevig2022,Si2010,Cao2018}. Narrow electronic bands can originate from two distinct pathways. In strongly correlated electron systems (large $U$), the Coulomb interactions localize the electronic orbitals on atomic sites with little or no overlap, resulting in extremely heavy quasiparticles or Mott insulating states. In these systems, the narrowing bandwidth is due to the suppressed overlap of the electronic wavefunctions, and the electronic states have suppressed spectral weight not visible to angle-resolved photoemission spectroscopy (ARPES), such as in the Mott insulators or localized $f$ orbitals in Kondo systems~\cite{Kirchner2020,Lee2006}. These large-$U$ flat band systems are exemplified by the heavy-fermion compounds, cuprates and iron-based superconductors~\cite{Checkelsky2024-yn}.

\begin{figure*}[]
\includegraphics[width=\textwidth]{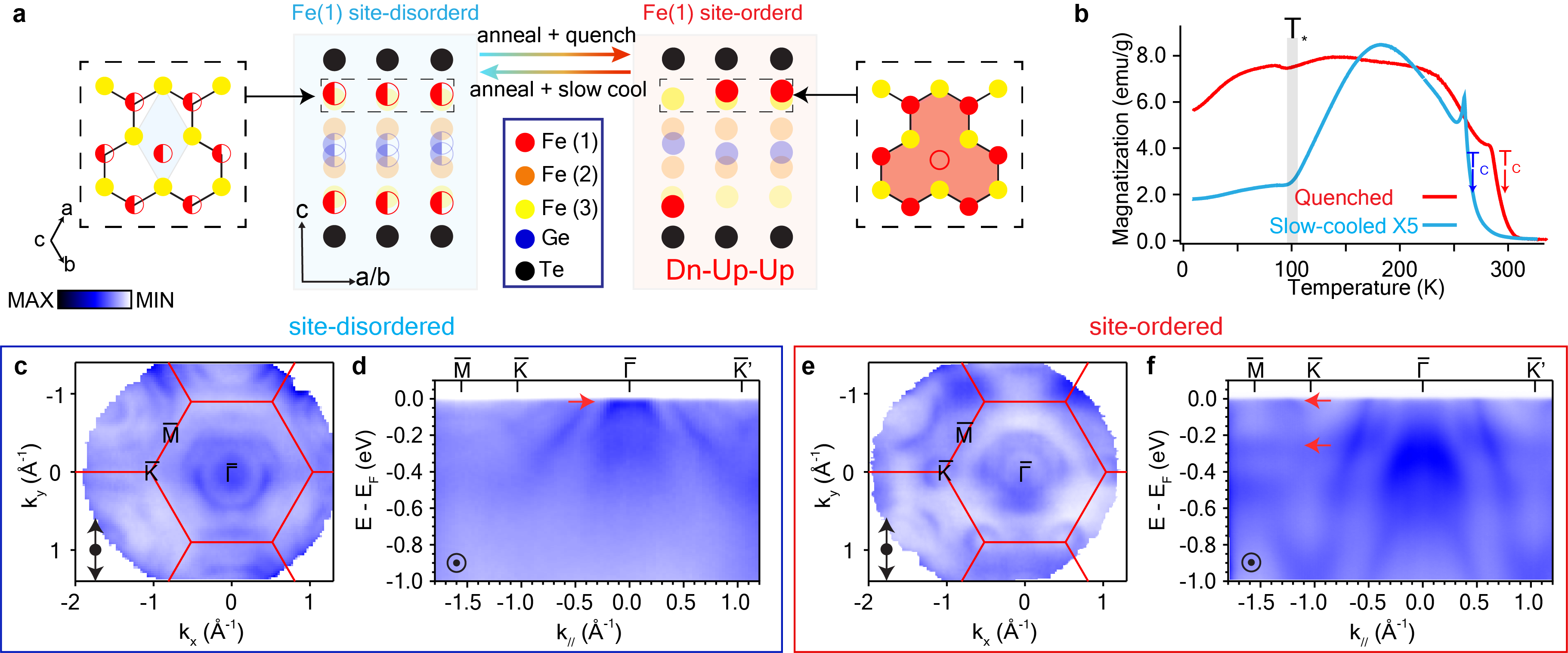} 
\caption{\textbf{Crystal structure, magnetization, and electronic structures of the site-disordered and site-ordered phases in Fe$_{5-x}$GeTe$_2$} (a) Lattice structure of a single vdW slab is contrasted between the site-disordered phase and site-ordered phase. The dashed boxes show the in-plane crystal structures at the indicated location within the vdW slab. Half-filled atoms indicate the 50\% occupied Fe(1) sites. Solid atoms indicate fully occupied sites. (b) Magnetization curves of a quenched crystal and a slow-cooled crystal. The quenched crystal is dominated by the site-ordered phase and slow-cooled crystal is dominated by site-disordered phase. (c) Measured Fermi surfaces and (d) dispersions along high symmetry direction for the site-disordered phase. (e)-(f) Same as (c)-(d) except measured for the site-ordered phase. The arrows in (d) and (f) indicate the flat bands. All data from (c) to (f) were taken with 114 eV photons at 15 K. Light polarization directions are as indicated.
%Note that in e the flat band is slight above \ef~thus only some intensity is observed. 
}
\label{fig:fig1}
\end{figure*}
Alternatively, flat bands can emerge due to the quantum destructive interference of geometrically frustrated lattices~\cite{Regnault2022,Calugaru2022,Yin2018,Kang2020,Kang2020_2}. In this case, the electronic wavefunctions are rather extended, and the destructive interference of the electronic wavefunction quenches the kinetic energy (small $t$), leading to flat bands in momentum space with suppressed bandwidth but full quasiparticle coherence. These flat bands are intrinsically topologically nontrivial as they obstruct a real-space description~\cite{HaoyuHu2023,Regnault2022}. Such flat bands, when tuned to the chemical potential, can lead to magnetism, fractional anomalous quantum Hall effects, or unconventional superconductivity~\cite{Calugaru2022,Regnault2022,Soumyanarayanan2016,Wang2017_Theory,Yin2018,Tasaki1992,Wu2007,Daniel2018,Tang2011,Sun2011,Neupert2011,Kirchner2020,Vergniory2019,Xu2020,Schindler2018,Barry2016,F3Zhang2018}. These small-$t$ flat band systems are exemplified by 2D kagome lattice materials including FeSn, Ni$_3$I, CsCr$_3$Sb$_5$, and 3D pyrochlore lattice compounds including Laves phases such as CeRu$_2$ and CaNi$_2$, and spinel compounds CuV$_2$S$_4$ and LiV$_2$O$_4$~\cite{Ren2024Persistent,Ye2024Hopping,Liu2024Superconductivity,Guo2024Ubiquitous,Li2025Electron,Peng2024Flat,Huang2024Observation,Wakefield2023Three,Huang2024Non}. 
In addition, small-$t$ flat band systems like Lieb lattice, Split lattice, Dice lattice have also been proposed extensively but have not been experimentally realized~\cite{Regnault2022}. Another type of destructive interference flat band system are the bipartite lattices, which are lattices with two different sublattices A and B with the nearest neighbor hopping between the two sublattices.  When the two sublattices have different number of orbitals, N$_A$ and N$_B$ per unit cell, due to the rank nullity theorem, N$_A$ - N$_B$ topological zero modes (or flat bands) would be guaranteed at the zero energy by destructive interference~\cite{Calugaru2022}. So far, experimental realization of bipartite crystalline materials have largely been restricted to optical, acoustic and superconducting circuits~\cite{Wirth2010Orbital,Zhu2022Observation,Marcos2014Lattice}. Only very recently, flat bands associated with bipartite lattice near the Fermi level have been experimentally reported in \fgt~\cite{wu2023reversible}.

\begin{figure*}[]
\includegraphics[width=\textwidth]{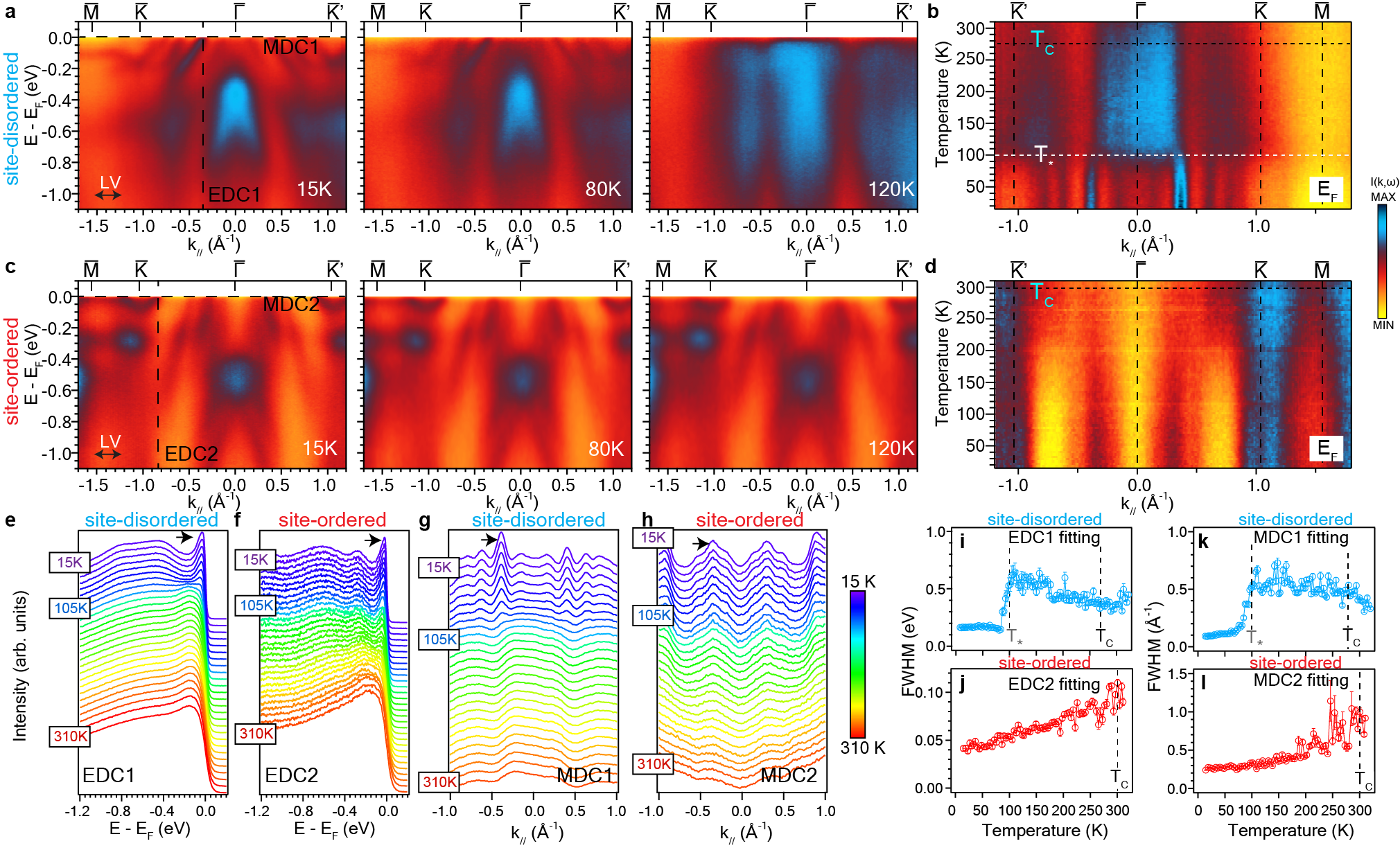} 
\caption{\textbf{Coherence-incoherence transition in the site-disordered phase.} (a) Band dispersions along the high symmetry direction measured in the site-disordered phase at 15 K, 80 K, and 120 K. (b) Temperature evolution of the momentum distribution curve (MDC) taken at the Fermi level in the site-disordered phase. (c)-(d) The same as (a)-(b) but measured in the site-ordered phase. (e)-(f) Temperature dependence of energy distribution curves (EDCs) for the two phases. The momentum location of the EDCs are labeled in (a) and (c). (g) Temperature-dependence of MDCs taken at \ef~in the site-disordered phase. (h) Temperature-dependence of MDCs taken at \ef~in the site-ordered phase.(i)-(j) Full width at half maximum (FWHM) of the EDC fitted peaks as a function of temperature. The fitted peaks are marked with black arrows in (e) and (f). (k)-(l) FWHM of the MDC-fitted peaks as a function of temperature. The fitted peaks are marked with black arrows in (g) and (h). The data from (a) to (b) were taken with 114 eV photons while the data from (c) to (d) were taken with 132 eV photons.
}
\label{fig:fig2}
\end{figure*}

Despite the distinct origin and properties of the two types of flat bands, analogies have been drawn in the theoretical treatment of the two types of flat bands~\cite{Calugaru2022,Regnault2022,Checkelsky2024-yn}. The coexistence of heavy fermions and itinerant carriers in AB-stacked transition metal dichalcogenide moiré bilayers and asymmetric bilayer graphene and hBN heterostructures~\cite{Zhao2023-wu, Vano2021-ug}, for example, have prompted analogies to Kondo lattices where localized moments (f-electrons) interact with itinerant carriers (s, p, d electrons), mapped to the topological flat bands intersecting with dispersive bands~\cite{Checkelsky2024-yn,Calugaru2022,Regnault2022,Kirchner2020}. Equally similar is the problem of orbital-selective Mott phase for multi-orbital d-electron systems, where selective orbital(s) become localized due to Coulomb interactions in the presence of other orbitals that remain metallic, as in the case of the iron-chalcogenides~\cite{Fernandes2022-yn,Yi2017-km,Yi2019,Huang2022-pw}. This analogy between the topological flat bands and those induced from strong correlations is meritorious in that a number of properties indeed have been found to be common between two types of material systems, including non-Fermi liquid behavior and strange metallicity. However, there has not been any direct comparison of the spectral properties of these two types of flat bands within one system, which is important for the understanding of the role of the two types of flat bands in emergent electronic orders.

The variations of the different underlying atomic or molecular environments prevent a side by side comparison of how their fundamentally distinct natures affect their behaviors in emergent electronic orders. 
Here, in addition to the previously reported geometrically frustrated flat bands in site-ordered van der Waals (vdW) \fgt~\cite{wu2023reversible}, we report the discovery of a correlation-induced flat band in the site-disordered \fgt, hence realizing two distinct types of flat bands in the same material system.  
Importantly, we observe characteristically distinct evolution of the two types of flat bands as a function of temperature through the respective ferromagnetic transitions. The flat band in the site-ordered phase displays a spin-splitting below the Curie temperature, while the flat band in the site-disordered phase is observed to gradually lose quasiparticle coherence with raised temperature. The former is compatible with the geometric frustration origin of its nature (small $t$) while the latter is compatible with the prototypical behavior of strong electron correlation effects (large $U$). In addition, we discuss the connection of these types of behaviors to the magnetic orders in the two phases. The contrast of the spectral evolution of the two types of flat bands in the same material system provides a direct visualization of the distinct mechanisms by which flat bands are generated in quantum materials, with distinctive consequences on their participation in the emergent orders. 

\section{Two flat bands of distinct origins}
Fe$_{5}$GeTe$_2$ belongs to a family of metallic ferromagnet, displaying an above-room-temperature Curie temperature ($T_\mathrm{C}$) ranging from 270 K to 330 K~\cite{F5_sky_Brian2023,F5_sky_Fujita2022,Li2020_F5,Zhang2020_F5,May2019_F5PRM,Li2021_F5ARPES,Wu2021_F5arpes,Huang2022_F5ARPES,Ly2021_F5ordering,May2019_F5ACSNANO,Ribeiro2022_F5MBE,Gao2020_F5domainwall,Chenxiang2022_F5doped,wu2023reversible,Wu2024-as}. A unique aspect of Fe$_{5}$GeTe$_2$ that distinguishes it from other Fe$_n$GeTe$_2$ (n=3 to 4) materials is a Fe site vacancy that is tunable~\cite{wu2023reversible,May2019_F5PRM,May2019_F5ACSNANO,Ly2021_F5ordering}.
The unit cell of Fe$_{5}$GeTe$_2$ consists of three ABC-stacked layers of vdW slabs. Within each slab, Fe and Ge sites are sandwiched by Te layers (Fig.~\ref{fig:fig1}a). Three distinct Fe sites are identified as Fe(1), Fe(2) and Fe(3), each appearing as an inversion symmetric pair about the inversion center at the Ge sites. As has been reported previously~\cite{May2019_F5PRM,May2019_F5ACSNANO}, stoichiometric Fe$_5$GeTe$_2$ structure has fully occupied Fe(2) and Fe(3) sites but each up-down inversion-symmetric pair of Fe(1) sites are half occupied. By slowly cooling the crystals from above a characteristic temperature, $T_\mathrm{HT}$, the Fe(1) sites are randomly occupied, leading to the site-disordered phase. 
%This lattice structure is rhombohedral in the space group $R\overline{3}m$, No. 166. 
By quenching the crystals from above $T_\mathrm{HT}$, the occupancy of the Fe(1) sites prefers to form an up-down-down (UDD) or down-up-up (DUU) pattern, resulting in an in-plane clover superstructure, as shown in Fig. 1a~\cite{Ly2021_F5ordering,May2019_F5ACSNANO,May2019_F5PRM,wu2023reversible}. 
While the two phases have the same stoichiometry, their physical properties are known to be distinct. In the magnetic susceptibility (Fig.~\ref{fig:fig1}b), for a crystal quenched from above $T_\mathrm{HT}$ with a dominant domain population of the site-ordered phase, the ferromagnetic ordering temperature, $T_\mathrm{C}$, is around 300 K. In contrast, when a crystal is cooled slowly from above $T_\mathrm{HT}$ with a predominantly site-disordered domain population, the $T_\mathrm{C}$ is lowered to 270 K, with an additional transition at 100 K ($T_{*}$)~\cite{wu2023reversible,May2019_F5ACSNANO, May2019_F5PRM}.

We carried out high quality ARPES measurement as a function of temperature for the two phases achieved by the aforementioned thermal methods. We first show the Fermi surface and high symmetry band dispersions at the lowest temperature of 15 K. Figure~\ref{fig:fig1}c-f show the Fermi surface and band dispersions measured for the two types of samples under the same measurement geometry, light polarization and photon energy. The contrast is remarkable, as reported previously~\cite{wu2023reversible}.
We focus on the band dispersions along the high symmetry direction M-K-$\Gamma$-K' as juxtaposed in Figs. 1d and f. In general, the dispersion slopes for the site-disordered phase are smaller than those for the site-ordered phase, indicating stronger correlation effects. As reported previously~\cite{wu2023reversible}, the clover-lattice of the site-ordered phase is a bipartite lattice that gives rise to topological flat bands from the destructive interference (pointed to by arrows in Fig.~\ref{fig:fig1}f).
In the site-disordered phase, weak spectral intensity is observed near~\ef~ as indicated by the red arrow in Fig.~\ref{fig:fig1}d. As there is no geometric frustration in this crystal structure, the origin of the flat band must be due to a different origin from that in the site-ordered phase.
As magnetization measurements indicate dramatic differences of the two types of crystals, we next present the temperature evolution of the electronic structures to gain more insights. 

\section{Coherence-incoherence transition}
Figure~\ref{fig:fig2}a-d show the detailed temperature evolution of the electronic structure for the site-disordered and site-ordered phase in Fe$_{5-x}$GeTe$_2$ taken with linear vertical (LV) polarized photons to probe more bands. In the site-disordered phase, we observe several hole-like band dispersions centered at $\Gamma$. Near -0.18~eV, two bands cross at the K point, forming a topological degenerate point which is protected by the combination of the global inversion symmetry (P), rotational symmetries $C_{2y}$ and $C_{3z}$~\cite{wu2023reversible}. In the site-ordered phase, the degeneracy at K is lifted due to the broken inversion symmetry. Noteworthily, Figs.~\ref{fig:fig2}b and d show a continuous temperature-dependent measurement of the momentum distribution curve (MDC) at \ef~along the high symmetry cut shown in Figs.~\ref{fig:fig2}a,c for the site-disordered and site-ordered phases, respectively. 

Compared with the smooth evolution in the site-ordered phase (Fig.~\ref{fig:fig2}d), the electronic structure in the site-disordered phase shows a discontinuous transition at $T_*$ = 100 K (Fig.~\ref{fig:fig2}b). From the dispersions across this transition (Fig.~\ref{fig:fig2}a), a broadening of all the bands above $T_*$ is clearly observed. 
The temperature evolution can also be visualized from the stack of the MDCs shown for the site-disordered phase (Fig.~\ref{fig:fig2}g) and the site-ordered phase (Fig.~\ref{fig:fig2}h). Equally clear are the trends from the energy distribution curves (EDCs) chosen at \kf~where the bands cross the \ef~from both phases as shown in Figs.~\ref{fig:fig2}e-f. 
By fitting the peaks from the stacked EDCs and MDCs, we can extract the full width at half maximum (FWHM) as a function of temperature. The trends from both EDCs and MDCs confirm a sudden broadening at $T_*$ for the site-disordered phase (Figs.~\ref{fig:fig2}i,k), indicating a coherence-incoherence transition where the quasiparticle suddenly becomes incoherent and thus ill-defined. In contrast, in the site-ordered phase, the temperature dependent quasiparticle energy and momentum widths evolves smoothly across $T_*$.
Previous Mossbauer results suggest that the anomaly in the magnetization curve at $T_*$ is from the onset of Fe(1) moment fluctuation~\cite{May2019_F5ACSNANO}. This would be compatible with the sudden increase in the quasiparticle linewidth corresponding to the site-disordered phase, due to the additional channel of electron-spin scattering. In contrast, the lack of any anomaly at $T_*$ in the site-ordered phase suggests that the coherence-incoherence transition is suppressed by the site-ordering. 

\begin{figure*}[]
\includegraphics[width=0.8\textwidth]{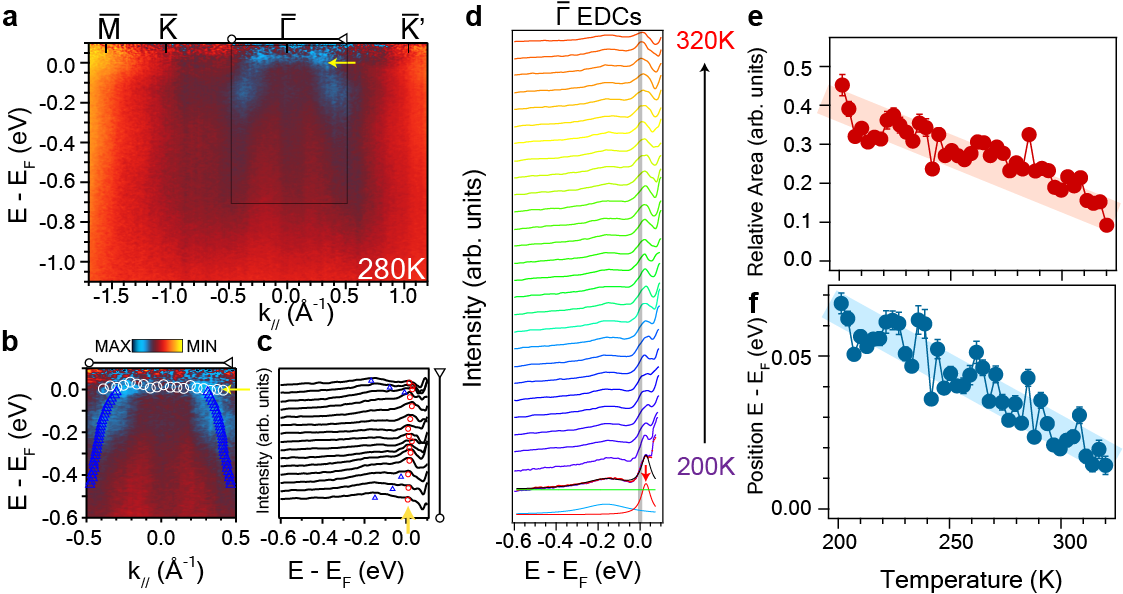} 
\caption{\textbf{Flat band in the site-disordered phase} (a) The band dispersions for the site-disordered phase measured at 280 K. The flat band marked by yellow arrow is hybridized with two dispersive bands. (b) Zoom-in view of the flat band and the dispersive bands. The white and blue markers are the location of the bands from the fittings. (c) shows the EDCs within the same region. The blue dots and red dots are fitted dispersive and flat bands positions. The yellow arrow points the flat band. (d) The EDCs that are taken at the $\Gamma$ point measured from 200 K to 320 K. The black line is the fitting where the flat band is fitted by the red Lorentzian peak. The red arrow points to the peak associated with the flat band. (e) The spectral weight of the fitted peak corresponding to the flat band as a function of temperature. (f) The fitted flat band position as a function of temperature. The data were taken with 114 eV photons.
}
\label{fig:fig3}
\end{figure*}
\section{Electron correlation-induced flat bands in site-disordered phase}
Having discussed the unique coherence-incoherence transition tuned by site ordering, we next present the second key observation--flat bands with distinct behaviors in the two phases. 
Upon examining the spectra near \ef~for the site-disordered phase, we noticed spectral weight in the near-$\Gamma$ region (Fig.~\ref{fig:fig1}d). After dividing by the Fermi-Dirac distribution convolved with the instrumental resolution, we observe a flat band located slightly above \ef~in the site-disordered phase (Fig.~\ref{fig:fig3}a).

Due to the Fermi-Dirac function, we can only observe this feature with sufficient thermal population at above 200K.

The flat band is observed to be hybridized with two dispersive bands near $\Gamma$.The fitted dispersions of the flat band and two dispersive bands measured at 280 K are shown in Fig. 3b,c (see SM for fitting details). The presence of this flat band is further confirmed from a different sample with a slightly higher Fe content but also in the site-disordered phase, where this flat band is located slightly below \ef~(see Fig. S3 in SM). As Density-Functional Theory calculations do not produce any flat bands near the $\Gamma$ point in Fe$_5$GeTe$_2$~\cite{wu2023reversible}, this suggests that this flat band may originate from correlation effects that strongly renormalizes it to the vicinity of \ef.

Furthermore, the spectral evolution of the flat band with temperature can be tracked by the EDC at $\Gamma$ after dividing by the Fermi-Dirac function, as shown in Fig.~\ref{fig:fig3}d. A peak can be observed near \ef~at all temperatures shown. The spectral weight of the flat band can be extracted from the peak area (Fig. 3e) while the location in energy from its position (Fig. 3f) (see SM for fitting details). We can clearly observe that the spectral weight of the flat band decreases with temperature, extrapolated to disappear at a temperature higher than $T_\mathrm{C}$. Simultaneously, the flat band shifts towards \ef~with increasing temperature. These behaviors are reminiscent of the behavior of correlation-driven flat bands, such as the case of orbital-selective Mott transition in the iron chalcogenides~\cite{Huang2022Correlation,Yi2015-cp,Yi2013-zy,Sobota2021Angle}, where bands of the selective orbital loses quasiparticle coherence and spectral weight as temperature is raised while its bandwidth diminishes with a renormalization that pushes it towards \ef~as the orbital approaches the Mott limit. In the present case of the site-disordered phase, this temperature scale appears to be above \tc.

\begin{figure*}[]
\includegraphics[width=0.8\textwidth]{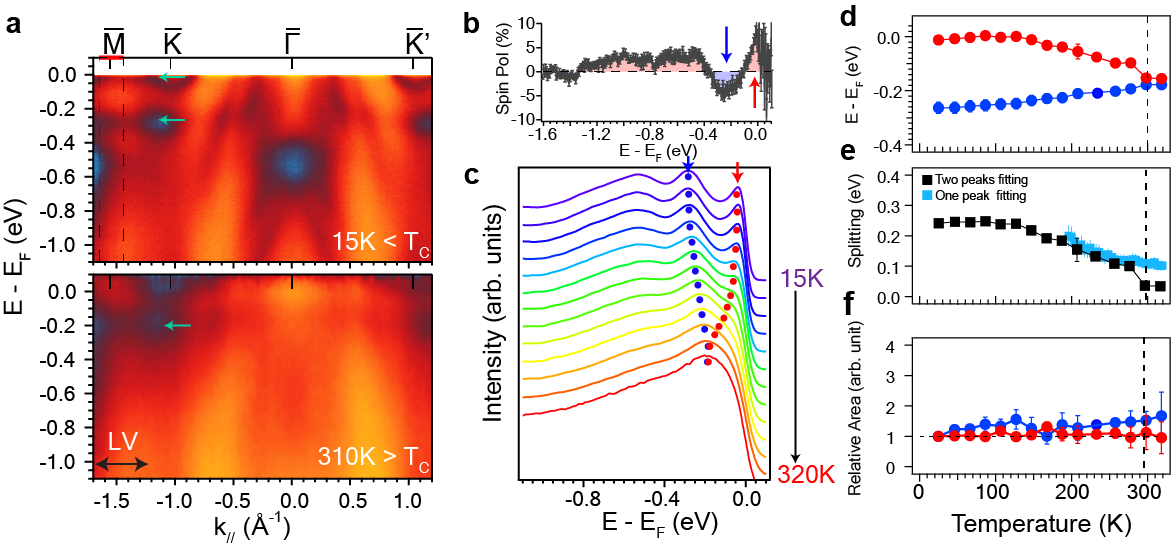} 
\caption{\textbf{Flat bands in the site-ordered phase} 
(a) The band dispersions for site-ordered phase taken at 15 K and 310 K. The arrows mark the two flat bands. (b) Spin polarization generated from spin-resolved EDCs taken at 15 K (see SM). (c) Temperature-dependent EDCs taken at M integrated between the dotted lines in (a). The two peaks marked by the red and blue dots are the two flat bands closest to \ef. (d)-(f) Fitted flat band positions, magnitude of the flat band splitting and spectral weight as a function of temperature, respectively. The ARPES data were taken with 132 eV photons. The spin-resolved ARPES data in (b) were taken with 66 eV photons.
}
\label{fig:fig4}
\end{figure*}

\section{Geometric frustration-induced flat bands in site-ordered phase}
In contrast to the site-disordered phase, the flat bands observed in the site-ordered phase have been discussed to be associated with the clover bipartite lattice~\cite{wu2023reversible}. We present the summary of the temperature evolution of these flat bands in Fig.~\ref{fig:fig4}. From the EDCs integrated near the M point as marked in Fig.~\ref{fig:fig4}a, we can track the flat bands from the peaks. Notably, the flat band closest to \ef~ and -0.2 eV (both marked in blue) move closer to each other as the temperature is raised, eventually substantially merge at $T_\mathrm{C}$. We can track the shift of the two bands by fitting the peak positions corresponding to the two flat bands (Fig.~\ref{fig:fig4}d). The two-peak fitting shows that the splitting largely disappears above $T_\mathrm{C}$ (see SM for fitting details). An independent method to check for the merging of the two bands can be carried out in a one-peak fit close to $T_\mathrm{C}$. As the two bands become close in energy, the FWHM of a single-peak fit could serve as a proxy of their temperature-dependent separation (see SM). The result of such a fit is shown in Fig.~\ref{fig:fig4}e, where a decreasing trend is clearly seen across $T_\mathrm{C}$, with the residual value above \tc~the upper bound for any residual splitting of the two bands overestimated by the linewidth of a single flat band. Hence while we cannot definitively distinguish the scenarios that the splitting between the two bands completely disappears above $T_\mathrm{C}$ or that a small splitting remains, we can confidently conclude that a substantial portion of the splitting at least disappears across the transition.

Moreover, the two-peak fitting also shows that the spectral weight of the two flat bands does not decrease as a function of temperature, in contrast to the flat band in the site-disordered phase. In addition, we have carried out spin-resolved ARPES measurements of the dispersions in the site-ordered phase. From the spin polarization plot (Fig.~\ref{fig:fig4}b), we find that the spin texture flips sign between the peak near \ef~and the peak near -0.2 eV, suggesting that the two flat bands are of the opposite spin (see SM for details). 

Taking all of the observations together, we observe that the flat bands in the site-ordered phase retain their quasiparticle coherence as a function of temperature and split into bands of opposite spins below $T_\mathrm{C}$. Such kind of spin-splitting below the ferromagnetic ordering temperature near the Fermi level is prototypical of itinerant magnetism and is rarely observed~\cite{Huang2021_R}. In fact, it is interesting to point out that amongst all of the known Fe-based vdW metallic ferromagnets, there are no such mean-field like spin splitting of the bands observed that onset at $T_\mathrm{C}$ except this pair of flat bands in the site-ordered phase Fe$_{5-x}$GeTe$_2$~\cite{F3Zhang2018,F3_yulinARPES,Wu2023-be,wu2023reversible}. In all of the other reported Fe$_{n}$GeTe$_2$ compounds even including the site-disordered phase, the lack of strong band evolution across $T_\mathrm{C}$ indicated the dominant presence of local moments. The behavior of the splitting of this particular pair of flat bands in the site-ordered phase contrasts that observed for the correlation-driven flat band in the site-disordered phase, and suggests the special nature of the origin of these flat bands in the site-ordered phase. In this case, we speculate that it is due to the geometric frustration associated with the clover bipartite lattice that preserves its quasiparticle coherence and itinerancy.
%We also compared the dispersive bands changing at 15 K and 200 K which is also plotted in Extended data figure 8. Instead of dramatic changes with k$_F$ and v$_F$ at those two temperatures in site-disordered phase, the dispersive bands show barely changes of the k$_F$ and v$_F$ that the k$_F$ is slightly becoming larger from 15 K to 200 K and v$_F$ shows almost no change in the site-ordered phase. The deviation behaviors demonstrate the site-ordered phase is away from Kondo lattice or correlation driven regime~\cite{Kirchner2020}

\section{Effect on magnetic order}
\begin{figure}[]
\includegraphics[width=0.5\textwidth]{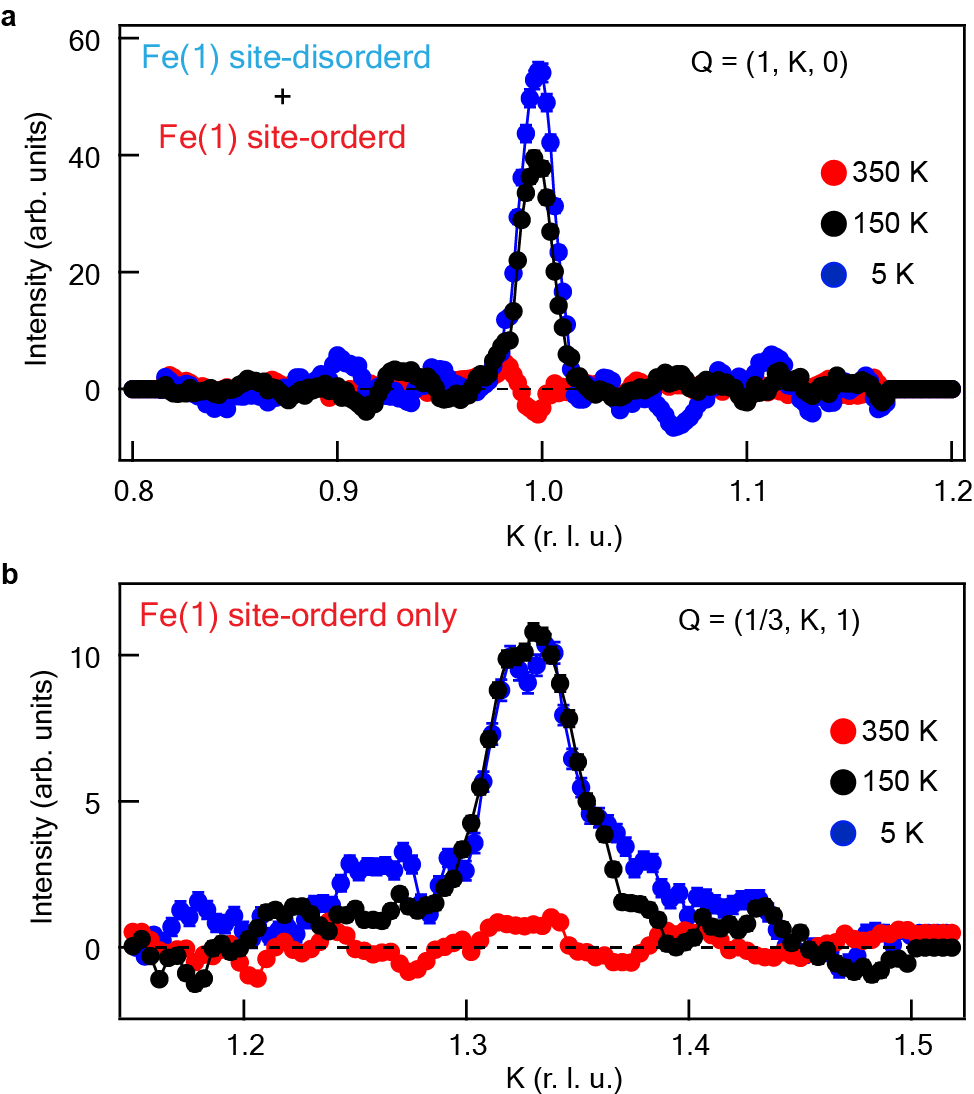} 
\caption{\textbf{Polarized neutron scattering in Fe$_5$GeTe$_2$} 
(a) Temperature-dependent polarized neutron scattering difference at Q = (1, K, 0). The magnetic signal at Q = (1, 1, 0) includes signals from both site-disordered and site-ordered regions. The data were integrated along the L direction within the window L $\in$ [-0.5, 0.5]. (b) Temperature-dependent polarized neutron scattering difference at Q = (1/3, K, 1). The magnetic signal at Q = (1/3, 4/3, 1) only has contribution from the site-ordered phase. The data were integrated along the L direction within the window L $\in$ [0, 2]. The difference in scattering signals from spin-up and spin-down neutrons for purely nuclear peaks should be zero, i.e., non-magnetic, such as the data at 350 K. The error bars are propagated from the statistical errors of 1 standard deviation.
}
\label{fig:fig5}
\end{figure}

Finally, we discuss the consequences of the site-ordering by comparing the magnetism in the two phases of Fe$_{5-x}$GeTe$_2$. As the lack of the coherence-incoherence transition in the site-ordered phase suggests that the magnetism is affected by the site-ordering, we carried out single-crystal unpolarized neutron diffraction measurement to directly probe the magnetism. As Fe$_{5-x}$GeTe$_2$ is ferromagnetic, magnetic peaks would appear on the nuclear peaks and can be tracked via changing temperature. Our measurement was carried out on a single crystal containing phase separated domains of the site-ordered and site-disordered phases (Fig. S9).
To isolate the signal of the mixed phases, we can track the superstructure peaks at fractional wavevectors corresponding to Q = ($\frac{1}{3}$, $\frac{1}{3}$, L), which can only correspond to the site-ordered regions. The remaining Bragg peaks contain signals from both site-ordered and site-disordered domains. 
%The superstructure peaks are only from the site-ordered phase and much broader along the L-direction than the Bragg peaks, suggesting a weak correlation between different layers. 
We note that as a function of temperature, no significant transition is observed for the ($\frac{1}{3}$, $\frac{1}{3}$, L) superstructure peaks at T$_{*}\sim$~100 K, consistent with the magnetization measurement on the quenched crystals with dominant site-ordered domain population, while an abrupt increase of peak intensities is observed at $T_*\sim$~100 K for peaks located at integer HKL values. To further confirm the magnetic orders and their evolution, we selected three temperatures of 6 K, 150 K, and 350 K to measure the corresponding peaks with polarized neutrons. At each temperature, selected peaks of Q = ($\frac{1}{3}$, $\frac{4}{3}$, 1) and (1, 1, 0) were measured with spin-up and spin-down neutrons, respectively. The difference of the spin-up and spin-down diffraction data are plotted in Fig.~\ref{fig:fig5}. Observable peaks in the difference plots confirm the conclusions drawn from the unpolarized measurement, namely that the site-ordered phase has no additional transition at T$_{*}$ as seen from the constant peak height at 6K and 150K for Q = ($\frac{1}{3}$, $\frac{4}{3}$, 1). In contrast, the peak at (1, 1, 0) has additional enhancement at 6K, which must come from the site-disordered region. The polarized neutron measurements confirm the magnetic order in both site-ordered and site-disordered phases, and demonstrates clearly that T$_{*}\sim$~100 K is only associated with the site-disordered phase.

Taking this observation together with the previous Mössbauer results we can conclude that moments on the Fe(1) sites remain fluctuating down to $T_{*}$ at which point they become ordered, but this only occurs when they are randomly distributed in the site-disordered phase~\cite{May2019_F5ACSNANO,Chen2023Thermal}. In contrast, for the site-ordered phase, the lack of moment enhancement at 100 K suggests that the Fe(1) moments are ordered together with the rest of the Fe moments at the higher $T_C$ of 300 K~\cite{Chen2023Thermal}. This is clearly compatible with the ARPES observation of a coherent-incoherent transition for the site-disordered phase and the lack of such a transition for the site-ordered phase as the fluctuating Fe(1) moments provide a spin scattering channel above $T_*$ for the site-disordered phase. Evidence of this can also be found in electrical transport measurements for the two types of crystals (see SM). Hence even though the two phases have the same average stoichiometry and local structure when considering a single Fe(1) site, the formation of the Fe(1) site-order is the key in dictating how the moments on these sites order.

\section{Discussion}
\begin{figure*}[]
\includegraphics[width=\textwidth]{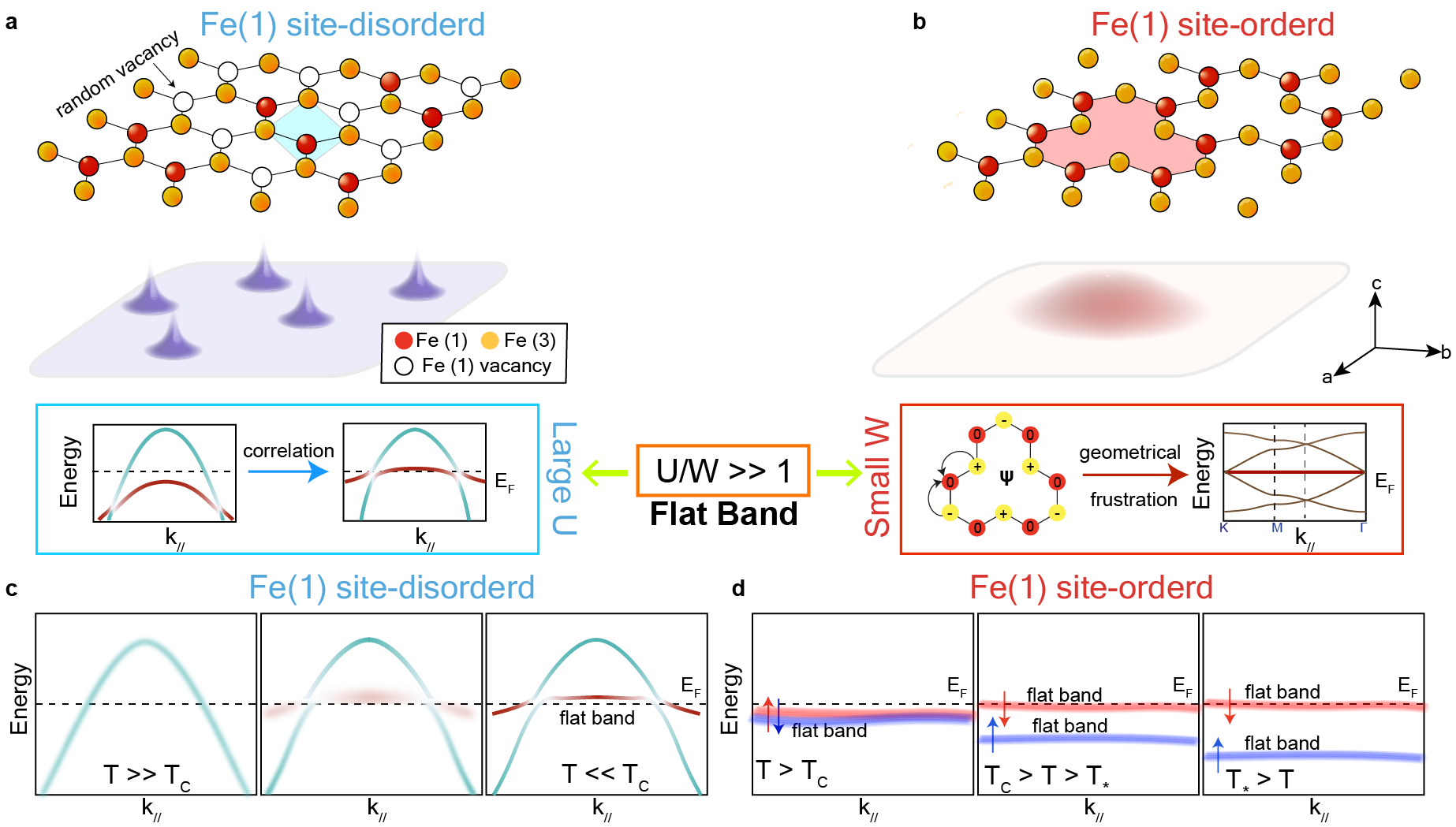} 
\caption{\textbf{Dichotomy of flat bands in Fe$_5$GeTe$_2$} 
(a) Schematic of the Fe(1)-Fe(3) plane in site-disordered phase. In the correlation-driven flat band, the Wannier functions associated with the flat bands are localized on the atomic sites. (b) Schematic of the Fe(1)-Fe(3) plane in site-ordered phase, where a clover lattice is formed. Here, the Bloch states are extended as the flat bands are induced by the quenching of kinetic energy from destructive interference. (c) Correlation-driven flat band observed in the site-disordered phase lose quasiparticle coherence with raised temperature. (d) Frustration-driven flat bands retain their quasiparticle coherence in the site-ordered phase but spin split below the ferromagnetic ordering temperature.
}
\label{fig:fig6}
\end{figure*}

To summarize, we have discovered a dichotomy of flat bands realized in the van der Waals ferromagnet Fe$_{5-x}$GeTe$_2$ that utilizes the formation of the Fe(1) site ordering as a tuning knob (Fig.~\ref{fig:fig6}). In one case, we have observed a phase with random occupation of Fe(1) sites a correlation-driven flat band that manifests the key signatures of electronic correlations--renormalized dispersions and gradual loss of spectral coherence as a function of temperature. In the other case, we observe the formation of the geometric lattice in the form of a fully coherent flat band that splits into spin up and spin down flat bands across $T_\mathrm{C}$.

We now summarize and advance the mechanisms to understand the dichotomy of two types of flat bands observed in the two phases and propose a possible explanation for the observed consequence on the material properties. In the site-ordered phase, the arrangements of the Fe(1) sites lead to the formation of the clover-like bipartite lattice that gives rise to the flat bands near \ef~with full quasiparticle coherence, which as the ferromagnetic order sets in become split into the spin up and spin down flat bands as observed in the low temperature phase. Here, the magnetic moments of all Fe sites are aligned ferromagnetically below $T_\mathrm{C}$. The strong ferromagnetic order can be understood by the induced Hund's-like direct exchange interaction between the nearest neighbor extended molecular orbital of the flat band. Since all the Fe atoms are fully polarized in the ferromagnetic phase, quantum fluctuations from interactions are not expected to be strong, as manifested in the lack of spectral weight suppression of the flat bands observed in the site-ordered phase. Conversely, the site-disordered phase is characterized by two characteristic temperatures, $T_{*}$ and $T_\mathrm{C}$. Different from the site-ordered phase, here, only Fe(2) and Fe(3) sites are ordered at $T_\mathrm{C}$, while Fe(1) sites remain fluctuating until below $T_{*}$. 
%Since it is no periodic potential in site-disordered phase due to the Fe(1) site disordering, the Fe(1) site is less likely close bound with Fe(2) and Fe(3) to order ferromagnetically together. 
According to the coherent potential approximation~\cite{Elliott1974-wc}, the average number of the nearest neighbor bonds connecting the Fe(1) and Fe(3) sites on the top is 1.5 in the site-disordered lattice, which is smaller than that in the site-ordered lattice with clover shape (which is 2). As a result, the kinetic energy in the site-disordered lattice is further reduced. Therefore, stronger quantum fluctuations are expected to be present in the Fe(1) orbitals, leading to a stronger manifestation of correlation effects. This is consistent with the observed overall smaller dispersion velocities in the site-disordered phase compared to the site-ordered phase (Fig.~\ref{fig:fig1}d,f), and also compatible with the observed gradual suppression of the spectral weight of the flat band in the site-disordered phase. 
%Furthermore, the weak correlated bipartite crystalline flat bands in the site-ordered phase exhibit a Stoner like mechanism which would further enhance the ferromagnetic order so the $T_\mathrm{C}$ is about 10\% higher in the quenched crystals. 

It is important to emphasize that the two types of flat bands exist in otherwise identical chemical environments hosted by the Fe$_5$GeTe$_2$ crystal structure including the other pairs of Fe(2), Fe(3), Ge and Te sites, with identical stoichiometry. Hence it is all the more remarkable that the spectral properties of these flat bands are so starkly contrasted. It is precisely this juxtaposition that allows us to directly ascribe these contrasting behaviors to the distinctive nature of the flat bands (large $U$ versus small $t$) and reveal the critical role played by the geometric frustration--something much less direct if considered apart (Fig.~\ref{fig:fig6}). Not only so, our work also clearly demonstrates the important role of geometric frustration on the magnetic properties, including the distinct ordering temperature of the Fe(1) sites.

We also discuss the spectral behavior of the flat bands in Fe$_5$GeTe$_2$ in the larger context of flat band systems. As many flat band systems have been reported, their nature and role in the physical properties paint a wide spectrum. Systems such as RbCo$_2$Se$_2$ exhibit flat bands naturally from DFT calculations, and hence do not require strong correlation nor geometric frustration and retain their full quasiparticle coherence~\cite{Huang2021_R}. Geometrically frustrated flat bands in kagome systems such as CsV$_3$Sb$_5$ and CoSn are far away from \ef~and hence do not play a dominant role in the physical properties~\cite{Wilson2024AV3Sb5,Ortiz2020,Kang2020_2,Liu2020CoSn}. Fe- and Mn-based kagome systems exemplified by FeGe, FeSn, and YMn$_6$Sn$_6$ do exhibit flat bands near \ef~in the non-magnetic calculations~\cite{Teng2022,Teng2023FeGe,Ren2024Persistent,Li2021YMn6Sn6}. However, the relatively large exchange splitting substantially remain across the magnetic ordering temperatures, and hence exhibit dominantly local moment physics~\cite{Ren2024Persistent}. Finally, kagome Ni$_3$I and pyrochlore CuV$_2$S$_4$ exhibit flat bands near \ef, but do not exhibit electronic orders driven by the flat bands, and hence are in a highly fluctuating regime with non-Fermi liquid transport behavior~\cite{Huang2024Non,Ye2024Hopping}. In this context, the geometrically frustrated flat bands in the site-ordered Fe$_5$GeTe$_2$ appear to be the first direct observation of spin splitting across a magnetic transition, indicating a previously unexplored regime of the larger phase diagram.

Lastly, what we have presented here may appear to be a unique case where correlation-driven flat bands and geometric frustration-driven flat bands can be effectively switched by the control of the sublattice site or vacancy order, in this case via the site-ordering of Fe(1) in Fe$_{5-x}$GeTe$_2$. However, fundamentally, as the origin of both types of flat bands is directly connected to the site-ordering---in one case the disorder-induced correlation enhancement and in the other the order-induced geometric frustration---this suggests that vacancy order can be explored as a more generic method for effectively controlling and tuning flat bands in quantum materials in a much wider context.

\section{Acknowledgments}
%------------------------------------------------------------------------------------------------------
This research used resources of the Advanced Light Source and the Stanford Synchrotron Radiation Lightsource, all U.S. Department Of Energy (DOE) Office of Science User Facilities under contract Nos. DE-AC02-05CH11231 and DE-AC02-76SF00515, respectively. The ARPES work from Rice is supported by the U.S. DOE grant No. DE-SC0021421, the Gordon and Betty Moore Foundation’s EPiQS Initiative through grant no. GBMF9470, and the Welch Foundation Grant No. C-2175 (M.Y.). YZ is partially supported by the Department of Defense, Air Force Office of Scientific Research under Grant No. FA9550-21-1-0343. The theory work at Rice is primarily supported by the U.S. DOE, BES, under Award No. DE-SC0018197 (Q.S.), by the AFOSR under Grant No. FA9550-21-1-0356 (Q.S.), and by the Robert A. Welch Foundation Grant No. C-1411 and the Vannevar Bush Faculty Fellowship ONR-VB N00014-23-1-2870 (Q.S.). The sample annealing process at Rice is supported by the U.S. DOE, BES under Grant No. DE-SC0012311 and the Robert A. Welch Foundation under grant no. C-1839 (P.D.).
The work at LBNL and UC Berkeley was funded by the U.S. Department of Energy, Office of Science, Office of Basic Energy Sciences, Materials Sciences and Engineering Division under Contract No. DE-AC02-05-CH11231 (Quantum Materials program KC2202). M. H. and D. L. acknowledge the support of the U.S. Department of Energy, Office of Science, Office of Basic Energy Sciences, Division of Material Sciences and Engineering, under contract DE-AC02-76SF00515. T.L.W. and Y.H. acknowledge the support from the Office of Naval Research under award N0014-23-1-2018.
A part of research using polarized neutrons at Oak Ridge National Laboratory (ORNL) was supported by the U.S. Department of Energy (DOE), Office of Science, Office of Basic Energy Sciences, Early Career Research Program Award KC0402020, under Contract DE-AC05-00OR22725. This research used resources at the High Flux Isotope Reactor, a DOE Office of Science User Facility operated by ORNL. Work at Los Alamos was carried out under the auspices of the U.S. Department of Energy (DOE) National Nuclear Security Administration (NNSA) under Contract No. 89233218CNA000001. It was supported by LANL LDRD Program ,and in part by Center for Integrated Nanotechnologies, a DOE BES user facility, in partnership with the LANL Institutional Computing Program for computational resources (B.G.J. \& J.-X.Z.).

\section{Methods}
\subsection{Crystal Synthesis}
Single crystals were grown via iodine-assisted chemical vapor transport following previous methods~\cite{May2019_F5PRM}. Fe powder, Ge pieces, and Te shot were weighed in the molar ratios 5:1:2, mixed, and placed within a quartz tube along with 2.509$ mg/cm^{3}$ of I$_{2}$ pieces. The tube was then sealed under low pressure Ar atmosphere and the sealed tubes were placed in a horizontal furnace with one end open to air to create a natural temperature gradient with the source material at the center of the furnace. The furnace was ramped to 750$^{\circ}$C over 12 hours, dwelled for two weeks, and then allowed to slowly cool back to room temperature for slow cooled samples. Fe$_{5-\delta}$GeTe$_2$ single crystals with plate-like morphology and mirror-like surfaces after cleaving grew at the cold end of the quartz tubes. The crystals always exhibits finite Fe deficiency. Single crystal x-ray diffraction refinements give a typical Fe deficiency of $\delta= \sim0.2$. As is known, for quenched Fe$_{5-\delta}$GeTe$_2$ crystals, there is a irreversible phase transition near 100 K upon the first cool-down~\cite{May2019_F5PRM}. To avoid complications, all our measurements presented for all techniques start after the first cool-down. For re-quenching a crystal, we sealed the crystals in a quartz ampoule, slowly ramped to 750 K and annealed for 2 hours then quenched in cold water.

\subsection{Magnetization}
Magnetization measurements were performed using the VSM option of the Quantum Design PPMS Dynacool. Samples were cleaved and shaped using a scalpel and mounted onto sample holders with a small amount of GE varnish. 

\subsection{ARPES and spin-resolved ARPES measurements}
Spin-integrated ARPES measurements were carried out at beamline 5-2 of the Stanford Synchrotron Radiation Lightsource using a DA30 electron analyzer. The energy and angular resolutions were set to 20 meV and 0.1$^\circ$, respectively. The spin-resolved ARPES measurements were carried out at beamline 10.0.1.2 of the Advanced Light Source with a DA30 electron analyzer and a VLEED spin detector. To align the ferromagnetic domains, a thin flake of neodymium permanent magnet was mounted below the crystal to provide a small magnetic field less than 50 Oe along the out-of-plane direction of the crystal. The samples were cleaved \textit{in-situ} at base temperature (between 15 and 20 K) and kept in ultra-high-vacuum with a base pressure lower than 5 $\times$ 10$^{-11}$ Torr during measurements. The Sherman function S value used for extracting the spin polarization is characterized as 0.24.

\subsection{Neutron scattering measurement}
Neutron diffraction measurements were performed at HB-3A DEMAND at High Flux Isotope Reactor (HFIR) at Oak Ridge National Lab~\cite{Neutron_1}. Single crystal neutron diffraction used unpolarized neutrons of 1.542 \AA~ from the bent monochromator Si (220)~\cite{Neutron_2}. The sample was mounted on the four-circle goniometer and cooled down to 5 K using a helium closed cycle refrigerator. The data reduction used ReTIA~\cite{Neutron_3}. The symmetry analysis used Bilbao Crystallography Server~\cite{Neutron_4}. The structure refinement used Fullprof Suite~\cite{neutron_5}. Neutrons of 2.5 \AA~ were used for half-polarized neutron diffraction measurements to validate the ferromagnetic order at both Fe(1) and other Fe sites. The crystal was loaded into a closed-cycle refrigerator with a permanent magnet set to provide a fixed field along the crystal c-axis for polarized neutron scattering measurement. The magnetic field is characterized as 0.53 T at 6 K and 0.69 T at 150 K due to temperature change. The polarized neutron data were normalized by the external field and corresponding magnetization curves. Neutrons were polarized by the S-bender supermirror polarizer and the polarization is larger than 95\%~\cite{Neutron_1}.

%------------------------------------------------------------------------------------------------------
%\section{Author Contributions}
%------------------------------------------------------------------------------------------------------
%The project was initiated and organized by MY. The single crystals were grown by CH, PM, YS, XX, and JC. The spin-integrated ARPES measurements and analyses were carried out by HW, JH, RJB and MY with the help of DL and MH. The spin-resolved ARPES measurements were carried out by HW, TW, SW, YH, and MY. The theory modeling was proposed and carried out by LC and QS. The neutron scattering experiments were done by YH, HC, and AM. The sample annealing and quenching process and characterization were carried out by PM, JC, YX, BG, HW and PD. The manuscript was written by HW and MY and contributed by all the authors.

%------------------------------------------------------------------------------------------------------

%------------------------------------------------------------------------------------------------------
\bibliographystyle{naturemag}
\bibliography{bib_new}
%------------------------------------------------------------------------------------------------------

\newpage

\end{document}